\font\ninerm=cmr9
\font\sevenrm=cmr7
\font\sixrm=cmr6
\font\fiverm=cmr5
\font\ninei=cmmi9
\font\sixi=cmmi6
\font\fivei=cmmi6
\font\ninesy=cmsy9
\font\sixsy=cmsy6
\font\fivesy=cmsy5
\font\tenex=cmex10
\font\nineit=cmti9
\font\ninesl=cmsl9
\font\ninett=cmtt9
\font\ninebf=cmbx9
\font\sixbf=cmbx6
\font\fivebf=cmbx5

\def\ninepoint{\def\rm{\fam0\ninerm}
  \textfont0=\ninerm \scriptfont0=\sixrm \scriptscriptfont0=\fiverm
  \textfont1=\ninei \scriptfont1=\sixi \scriptscriptfont0=\fivei
  \textfont2=\ninesy \scriptfont2=\sixsy \scriptscriptfont2=\fivesy
  \textfont3=\tenex \scriptfont3=\tenex \scriptscriptfont3=\tenex
  \textfont\itfam=\nineit  \def\it{\fam\itfam\nineit}%
  \textfont\slfam=\ninesl  \def\sl{\fam\slfam\ninesl}%
  \textfont\ttfam=\ninett  \def\tt{\fam\ttfam\ninett}%
  \textfont\bffam=\ninebf  \scriptfont\bffam=\sixbf
   \scriptscriptfont\bffam=\fivebf  \def\bf{\fam\bffam\ninebf}%
  \normalbaselineskip=11pt
  \setbox\strutbox=\hbox{\vrule height8pt depth3pt width0pt}%
  \let\sc=\sevenrm  \normalbaselines\rm}

\font\eightrm=cmr8
\font\sevenrm=cmr7
\font\sixrm=cmr6
\font\fiverm=cmr5
\font\eighti=cmmi8
\font\sixi=cmmi6
\font\fivei=cmmi6
\font\eightsy=cmsy8
\font\sixsy=cmsy6
\font\fivesy=cmsy5
\font\tenex=cmex10
\font\eightit=cmti8
\font\eightsl=cmsl8
\font\eighttt=cmtt8
\font\eightbf=cmbx8
\font\sixbf=cmbx6
\font\fivebf=cmbx5

\def\eightpoint{\def\rm{\fam0\eightrm}
  \textfont0=\eightrm \scriptfont0=\sixrm \scriptscriptfont0=\fiverm
  \textfont1=\eighti \scriptfont1=\sixi \scriptscriptfont0=\fivei
  \textfont2=\eightsy \scriptfont2=\sixsy \scriptscriptfont2=\fivesy
  \textfont3=\tenex \scriptfont3=\tenex \scriptscriptfont3=\tenex
  \textfont\itfam=\eightit  \def\it{\fam\itfam\eightit}%
  \textfont\slfam=\eightsl  \def\sl{\fam\slfam\eightsl}%
  \textfont\ttfam=\eighttt  \def\tt{\fam\ttfam\eighttt}%
  \textfont\bffam=\eightbf  \scriptfont\bffam=\sixbf
   \scriptscriptfont\bffam=\fivebf  \def\bf{\fam\bffam\eightbf}%
  \normalbaselineskip=9pt
  \setbox\strutbox=\hbox{\vrule height7pt depth2pt width0pt}%
  \let\sc=\sixrm  \normalbaselines\rm}

\catcode `\!=11
\catcode `\@=11

\let\!tacr=\\ 

\newdimen\LineThicknessUnit
\newdimen\StrutUnit
\newskip \InterColumnSpaceUnit
\newdimen\ColumnWidthUnit
\newdimen\KernUnit

\let\!taLTU=\LineThicknessUnit 
\let\!taCWU=\ColumnWidthUnit   
\let\!taKU =\KernUnit          

\newtoks\NormalTLTU
\newtoks\NormalTSU
\newtoks\NormalTICSU
\newtoks\NormalTCWU
\newtoks\NormalTKU

\NormalTLTU={1in \divide \LineThicknessUnit by 300 }
\NormalTSU ={\normalbaselineskip
  \divide \StrutUnit by 11 }  
\NormalTICSU={.5em plus 1fil minus .25em}  
\NormalTCWU ={.5em}
\NormalTKU  ={.5em}

\def\NormalTableUnits{%
  \LineThicknessUnit   =\the\NormalTLTU
  \StrutUnit           =\the\NormalTSU
  \InterColumnSpaceUnit=\the\NormalTICSU
  \ColumnWidthUnit     =\the\NormalTCWU
  \KernUnit            =\the\NormalTKU}

\NormalTableUnits

\newcount\LineThicknessFactor
\newcount\StrutHeightFactor
\newcount\StrutDepthFactor
\newcount\InterColumnSpaceFactor
\newcount\ColumnWidthFactor
\newcount\KernFactor
\newcount\VspaceFactor

\newcount\TracingKeys 
\newcount\TracingFormats  

\def\BeginTableParBox#1{%
  \vtop\bgroup
    \hsize=#1
    \normalbaselines
    \let~=\!ttTie
    \let\-=\!ttDH
    \the\EveryTableParBox}

\def\EndTableParBox{%
    \MakeStrut{0pt}{\StrutDepthFactor\StrutUnit}
  \egroup} 

\newtoks\EveryTableParBox
\EveryTableParBox={%
  \parindent=0pt
  \raggedright
  \rightskip=0pt plus 4em 
  \relax}

\newtoks\EveryTable
\newtoks\!taTableSpread

\newskip\LeftTabskip
\newskip\RightTabskip

\newcount\!taCountA
\newcount\!taColumnNumber
\newcount\!taRecursionLevel 

\newdimen\!taDimenA  
\newdimen\!taDimenB  
\newdimen\!taDimenC  
\newdimen\!taMinimumColumnWidth

\newtoks\!taToksA

\newtoks\!taPreamble
\newtoks\!taDataColumnTemplate
\newtoks\!taRuleColumnTemplate
\newtoks\!taOldRuleColumnTemplate
\newtoks\!taLeftGlue
\newtoks\!taRightGlue

\newskip\!taLastRegularTabskip

\newif\if!taDigit
\newif\if!taBeginFormat
\newif\if!taOnceOnlyTabskip

\def\TaBlE{%
  T\kern-.27em\lower.5ex\hbox{A}\kern-.18em B\kern-.1em
    \lower.5ex\hbox{L}\kern-.075em E}

{\catcode`\|=13 \catcode`\"=13
  \gdef\ActivateBarAndQuote{%
    \ifnum \catcode`\|=13
    \else
      \catcode`\|=13
      \def|{%
        \ifmmode
          \vert
        \else
          \char`\|
        \fi}%
    \fi
    \ifnum \catcode`\"=13
    \else
      \catcode`\"=13
      \def"{\char`\"}%
    \fi}}

{\catcode `\|=12 \catcode `\"=12

}

\def\!thMessage#1{\immediate\write16{#1}\ignorespaces}

\let\!thx=\expandafter

\def\!thGobble#1{}

\def\\{\let\!thSpaceToken= }\\

\def\!thHeight{height}
\def\!thDepth{depth}
\def\!thWidth{width}

\def\!thToksEdef#1=#2{%
  \edef\!ttemp{#2}%
  #1\!thx{\!ttemp}%
  \ignorespaces}

\def\!thStoreErrorMsg#1#2{%
  \toks0 =\!thx{\csname #2\endcsname}%
  \edef#1{\the\toks0 }}

\def\!thReadErrorMsg#1{%
  \!thx\!thx\!thx\!thGobble\!thx\string #1}

\def\!thError#1#2{%
  \begingroup
    \newlinechar=`\^^J%
    \edef\!ttemp{#2}%
    \errhelp=\!thx{\!ttemp}%
    \!thMessage{%
      ^^J\!thReadErrorMsg\!thErrorMsgA
      ^^J\!thReadErrorMsg\!thErrorMsgB}%
    \errmessage{#1}%
  \endgroup}

\!thStoreErrorMsg\!thErrorMsgA{%
  TABLE error; see manual for explanation.}
\!thStoreErrorMsg\!thErrorMsgB{%
  Type \space H <return> \space for immediate help.}

\def\!thGetReplacement#1#2{%
   \begingroup
     \!thMessage{#1}
     \endlinechar=-1
     \global\read16 to#2%
   \endgroup}

\def\!thLoop#1\repeat{%
  \def\!thIterate{%
    #1%
    \!thx \!thIterate
    \fi}%
  \!thIterate
  \let\!thIterate\relax}

\def\Smash{%
  \relax
  \ifmmode
    \expandafter\mathpalette
    \expandafter\!thDoMathVCS
  \else
    \expandafter\!thDoVCS
  \fi}

\def\!thDoVCS#1{%
  \setbox\z@\hbox{#1}%
  \!thFinishVCS}

\def\!thDoMathVCS#1#2{%
  \setbox\z@\hbox{$\m@th#1{#2}$}%
  \!thFinishVCS}

\def\!thFinishVCS{%
  \vbox to\z@{\vss\box\z@\vss}}

\def\Lower{%
  \def\!thSign{-}%
  \!tgGetValue\!thSetDimen}

\def\!thSetDimen{%
  \ifnum \!tgCode=1
    \ifx \!tgValue\empty
      \!taDimenA \StrutHeightFactor\StrutUnit
      \advance \!taDimenA \StrutDepthFactor\StrutUnit
      \divide \!taDimenA 2
    \else
      \!taDimenA \!tgValue\StrutUnit
    \fi
  \else
    \!taDimenA \!tgValue
  \fi
  \!taDimenA=\!thSign\!taDimenA\relax
  \ifmmode
    \expandafter\mathpalette
    \expandafter\!thDoMathRaise
  \else
    \expandafter\!thDoSimpleRaise
  \fi}

\def\!thDoSimpleRaise#1{%
  \setbox\z@\hbox{\raise \!taDimenA\hbox{#1}}%
  \!thFinishRaise} 

\def\!thDoMathRaise#1#2{%
  \setbox\z@\hbox{\raise \!taDimenA\hbox{$\m@th#1{#2}$}}%
  \!thFinishRaise}

\def\!thFinishRaise{%
  \ht\z@\z@
  \dp\z@\z@
  \box\z@}

\def\!thKernBack{%
  \kern -
  \ifnum \!tgCode=1
    \ifx \!tgValue\empty
      \the\KernFactor
    \else
      \!tgValue    
    \fi
    \KernUnit
  \else
    \!tgValue      
  \fi
  \ignorespaces}%

\def\Vspace{%
  \noalign
  \bgroup
  \!tgGetValue\!thVspace}

\def\!thVspace{%
  \vskip
    \ifnum \!tgCode=1
      \ifx \!tgValue\empty
        \the\VspaceFactor
      \else
        \!tgValue    
      \fi
      \StrutUnit
    \else
      \!tgValue      
    \fi
  \egroup} 

\def\BeginFormat{%
  \catcode`\|=12 
  \catcode`\"=12 
  \!taPreamble={}%
  \!taColumnNumber=0
  \skip0 =\InterColumnSpaceUnit
  \multiply\skip0 \InterColumnSpaceFactor
  \divide\skip0 2
  \!taRuleColumnTemplate=\!thx{%
    \!thx\tabskip\the\skip0 }%
  \!taLastRegularTabskip=\skip0
  \!taOnceOnlyTabskipfalse
  \!taBeginFormattrue 
  \def\!tfRowOfWidths{}
  \ReadFormatKeys}

\def\!tfSetWidth{%
  \ifx \!tfRowOfWidths \empty  
    \ifnum \!taColumnNumber>0  
      \begingroup              
         \!taCountA=1          
         \aftergroup \edef \aftergroup \!tfRowOfWidths \aftergroup {%
           \aftergroup &\aftergroup \omit
           \!thLoop
             \ifnum \!taCountA<\!taColumnNumber
             \advance\!taCountA 1
             \aftergroup \!tfAOAO
           \repeat
           \aftergroup }%
      \endgroup
    \fi
  \fi
  \ifx [\!ttemp 
    \!thx\!tfSetWidthText
  \else
    \!thx\!tfSetWidthValue
  \fi}

\def\!tfAOAO{%
  &\omit&\omit}

\def\!tfSetWidthText [#1]{
  \def\!tfWidthText{#1}%
  \ReadFormatKeys}

\def\!tfSetWidthValue{%
  \!taMinimumColumnWidth =
    \ifnum \!tgCode=1
      \ifx\!tgValue\empty 
        \ColumnWidthFactor
      \else
        \!tgValue
      \fi
      \ColumnWidthUnit
    \else
      \!tgValue
    \fi
  \def\!tfWidthText{}
  \ReadFormatKeys}

\def\!tfSetTabskip{%
  \ifnum \!tgCode=1
    \skip0 =\InterColumnSpaceUnit
    \multiply\skip0
      \ifx \!tgValue\empty
        \InterColumnSpaceFactor         
      \else
       \!tgValue                        
      \fi
  \else
    \skip0 =\!tgValue                   
  \fi
  \divide\skip0 by 2
  \ifnum\!taColumnNumber=0
    \!thToksEdef\!taRuleColumnTemplate={%
      \the\!taRuleColumnTemplate
      \tabskip \the\skip0 }
  \else
    \!thToksEdef\!taDataColumnTemplate={%
      \the\!taDataColumnTemplate
      \tabskip \the\skip0 }
  \fi
  \if!taOnceOnlyTabskip
  \else
    \!taLastRegularTabskip=\skip0 
  \fi                             
  \ReadFormatKeys}

\def\!tfSetVrule{%
  \!thToksEdef\!taRuleColumnTemplate={%
    \noexpand\hfil
    \noexpand\vrule
    \noexpand\!thWidth
    \ifnum \!tgCode=1
      \ifx \!tgValue\empty
        \the\LineThicknessFactor      
      \else
        \!tgValue                     
      \fi
      \!taLTU                         
    \else
      \!tgValue                       
    \fi
    ####%
    \noexpand\hfil
    \the\!taRuleColumnTemplate}       
  \!tfAdjoinPriorColumn}

\def\!tfSetAlternateVrule{%
  \afterassignment\!tfSetAlternateA
  \toks0 =}                           

\def\!tfSetAlternateA{%
  \!thToksEdef\!taRuleColumnTemplate={%
    \the\toks0 \the\!taRuleColumnTemplate} 
  \!tfAdjoinPriorColumn}

\def\!tfAdjoinPriorColumn{%
  \ifnum \!taColumnNumber=0
    \!taPreamble=\!taRuleColumnTemplate 
    \ifnum \TracingFormats>0
      \!tfShowRuleTemplate
    \fi
  \else
    \ifx\!tfRowOfWidths\empty  
    \else
      \!tfUpdateRowOfWidths
    \fi
    \!thToksEdef\!taDataColumnTemplate={%
      \the \!taLeftGlue
      \the \!taDataColumnTemplate
      \the \!taRightGlue}
    \ifnum \TracingFormats>0
      \!tfShowTemplates
    \fi
    \!thToksEdef\!taPreamble={%
      \the\!taPreamble
      &
      \the\!taDataColumnTemplate
      &
      \the\!taRuleColumnTemplate}
  \fi
  \advance \!taColumnNumber 1
  \if!taOnceOnlyTabskip
    \!thToksEdef\!taDataColumnTemplate={%
       ####\tabskip \the\!taLastRegularTabskip}
  \else
    \!taDataColumnTemplate{##}%
  \fi
  \!taRuleColumnTemplate{}
  \!taLeftGlue{\hfil}
  \!taRightGlue{\hfil}%
  \!taMinimumColumnWidth=0pt
  \def\!tfWidthText{}%
  \!taOnceOnlyTabskipfalse    
  \ReadFormatKeys}

\def\!tfUpdateRowOfWidths{%
  \ifx \!tfWidthText\empty
  \else 
    \!tfComputeMinColWidth
  \fi
  \edef\!tfRowOfWidths{%
    \!tfRowOfWidths
    &%
    \omit                                  
    \ifdim \!taMinimumColumnWidth>0pt
      \hskip \the\!taMinimumColumnWidth
    \fi
    &
    \omit}}                                

\def\!tfComputeMinColWidth{%
  \setbox0 =\vbox{%
    \ialign{
       \span\the\!taDataColumnTemplate\cr
       \!tfWidthText\cr}}%
  \!taMinimumColumnWidth=\wd0 }

\def\!tfShowRuleTemplate{%
  \!thMessage{}
  \!thMessage{TABLE FORMAT}
  \!thMessage{Column: Template}
  \!thMessage{%
    \space *c: ##\tabskip \the\LeftTabskip}
  \!taOldRuleColumnTemplate=\!taRuleColumnTemplate}

\def\!tfShowTemplates{%
  \!thMessage{%
    \space \space r: \the\!taOldRuleColumnTemplate}
  \!taOldRuleColumnTemplate=\!taRuleColumnTemplate
  \!thMessage{%
    \ifnum \!taColumnNumber<10
      \space
    \fi
    \the\!taColumnNumber c: \the\!taDataColumnTemplate}
  \ifdim\!taMinimumColumnWidth>0pt
    \!thMessage{%
      \space \space w: \the\!taMinimumColumnWidth}
  \fi}

\def\!tfFinishFormat{%
  \ifnum \TracingFormats>0
    \!thMessage{%
      \space \space r: \the\!taOldRuleColumnTemplate
        \tabskip \the\RightTabskip}%
    \!thMessage{%
      \space *c: ##\tabskip 0pt}
  \fi
  \ifnum \!taColumnNumber<2
    \!thError{%
      \ifnum \!taColumnNumber=0
        No
      \else
        Only 1
      \fi
      "|"}%
      {\!thReadErrorMsg\!tfTooFewBarsA
       ^^J\!thReadErrorMsg\!tfTooFewBarsB
       ^^J\!thReadErrorMsg\!tkFixIt}%
  \fi
  \!thToksEdef\!taPreamble={%
    ####\tabskip\LeftTabskip
    &
    \the\!taPreamble \tabskip\RightTabskip
    &
    ####\tabskip 0pt \cr}
  \ifnum \TracingFormats>1
    \!thMessage{Preamble=\the\!taPreamble}
  \fi
  \ifnum \TracingFormats>2
    \!thMessage{Row Of Widths="\!tfRowOfWidths"}
  \fi
  \!taBeginFormatfalse 
  \catcode`\|=13
  \catcode`\"=13
  \!ttDoHalign}

\!thStoreErrorMsg\!tfTooFewBarsA{%
  There must be at least 2 "|"'s (and/or "\string \|"'s)}
\!thStoreErrorMsg\!tfTooFewBarsB{%
  between \string\BeginFormat\space and \string\EndFormat\space (or ".").}

\def\ReFormat[{%
  \omit
  \!taDataColumnTemplate{##}%
  \!taLeftGlue{}%
  \!taRightGlue{}%
  \catcode`\|=12  
  \catcode`\"=12  
  \ReadFormatKeys}

\def\!tfEndReFormat{%
  \ifnum \TracingFormats>0
    \!thMessage{ReF:
       \the\!taLeftGlue
       \hbox{\the\!taDataColumnTemplate}
       \the\!taRightGlue}
  \fi
  \catcode`\|=13
  \catcode`\"=13
  \!tfReFormat}

\def\!tfReFormat#1{%
  \the \!taLeftGlue
  \vbox{%
    \ialign{%
      \span\the\!taDataColumnTemplate\cr
       #1\cr}}%
  \the \!taRightGlue}

\def\!tgGetValue#1{%
  \def\!tgReturn{#1}
  \futurelet\!ttemp\!tgCheckForParen}

\def\!tgCheckForParen{%
  \ifx\!ttemp (%
    \!thx \!tgDoParen
  \else
    \!thx \!tgCheckForSpace
  \fi}

\def\!tgDoParen(#1){%
  \def\!tgCode{2}%
  \def\!tgValue{#1}
  \!tgReturn}

\def\!tgCheckForSpace{%
  \def\!tgCode{1}%
  \def\!tgValue{}
  \ifx\!ttemp\!thSpaceToken
    \!thx \!tgReturn        
  \else
    \!thx \!tgCheckForDigit
  \fi}

\def\!tgCheckForDigit{%
  \!taDigitfalse
  \ifx 0\!ttemp
    \!taDigittrue
  \else
    \ifx 1\!ttemp
      \!taDigittrue
    \else
      \ifx 2\!ttemp
        \!taDigittrue
      \else
        \ifx 3\!ttemp
          \!taDigittrue
        \else
          \ifx 4\!ttemp
            \!taDigittrue
          \else
            \ifx 5\!ttemp
              \!taDigittrue
            \else
              \ifx 6\!ttemp
                \!taDigittrue
              \else
                \ifx 7\!ttemp
                  \!taDigittrue
                \else
                  \ifx 8\!ttemp
                    \!taDigittrue
                  \else
                    \ifx 9\!ttemp
                      \!taDigittrue
                    \fi
                  \fi
                \fi
              \fi
            \fi
          \fi
        \fi
      \fi
    \fi
  \fi
  \if!taDigit
    \!thx \!tgGetNumber
  \else
    \!thx \!tgReturn
  \fi}

\def\!tgGetNumber{%
  \afterassignment\!tgGetNumberA
  \!taCountA=}
\def\!tgGetNumberA{%
  \edef\!tgValue{\the\!taCountA}%
  \!tgReturn}

\def\!tgSetUpParBox{%
  \edef\!ttemp{%
    \noexpand \ReadFormatKeys
    b{\noexpand \BeginTableParBox{%
      \ifnum \!tgCode=1
        \ifx \!tgValue\empty
          \the\ColumnWidthFactor
        \else
          \!tgValue    
        \fi
        \!taCWU        
      \else
        \!tgValue      
      \fi}}}%
  \!ttemp
  a{\EndTableParBox}}

\def\!tgInsertKern{%
  \edef\!ttemp{%
    \kern
    \ifnum \!tgCode=1
      \ifx \!tgValue\empty
        \the\KernFactor
      \else
        \!tgValue    
      \fi
      \!taKU         
    \else
      \!tgValue      
    \fi}%
  \edef\!ttemp{%
    \noexpand\ReadFormatKeys
    \ifh@            
      b{\!ttemp}
    \fi
    \ifv@            
      a{\!ttemp}
    \fi}%
  \!ttemp}

\def\NewFormatKey#1{%
  \!thx\def\!thx\!ttempa\!thx{\string #1}%
  \!thx\def\!thx\!ttempb\!thx{\csname !tk<\!ttempa>\endcsname}%
  \ifnum \TracingKeys>0
    \!tkReportNewKey
  \fi
  \!thx\ifx \!ttempb \relax
    \!thx\!tkDefineKey
  \else
    \!thx\!tkRejectKey
  \fi}

\def\!tkReportNewKey{%
  \!taToksA\!thx{\!ttempa}%
  \!thMessage{NEW KEY: "\the\!taToksA"}}

\def\!tkDefineKey{%
  \!thx\def\!ttempb}%

\def\!tkRejectKey{%
    \!taToksA\!thx{\!ttempa}%
    \!thError{Key letter "\the\!taToksA" already used}
      {\!thReadErrorMsg\!tkFixIt}
    \def\!tkGarbage}%

\!thStoreErrorMsg\!tkFixIt{%
  You'd better type \space 'E' \space and fix your file.}

\def\ReadFormatKeys#1{%
  \!thx\def\!thx\!ttempa\!thx{\string #1}%
  \!thx\def\!thx\!ttempb\!thx{\csname !tk<\!ttempa>\endcsname}%
  \ifnum \TracingKeys>1
    \!tkReportKey
  \fi
  \!thx\ifx \!ttempb\relax
    \!thx\!tkReplaceKey
  \else
    \!thx\!ttempb
  \fi}

\def\!tkReportKey{%
  \!taToksA\!thx{\!ttempa}%
  \!thMessage{KEY: "\the\!taToksA"}}

\def\!tkReplaceKey{%
  \!taToksA\!thx{\!ttempa}%
  \!thError {Undefined format key "\the\!taToksA"}
    {\!thReadErrorMsg\!tkUndefined ^^J\!thReadErrorMsg\!tkBadKey}
  \!tkReplaceKeyA}

\def\!tkReplaceKeyA{%
  \!thGetReplacement{\!thReadErrorMsg\!tkReplace}\!tkReplacement
  \!thx\ReadFormatKeys\!tkReplacement}

\!thStoreErrorMsg\!tkUndefined{%
  The format key in " "'s on the next to top line is undefined.}
\!thStoreErrorMsg\!tkBadKey{%
  Type \space E \space to quit now, or
  \space<CR> \space and respond to next prompt.}
\!thStoreErrorMsg\!tkReplace{%
  Type \space<replacement key><CR> \space,
   or simply \space<CR> \space to skip offending key:}

\NewFormatKey b#1{%
  \!thx\!tkJoin\!thx{\the\!taDataColumnTemplate}{#1}%
  \ReadFormatKeys}

\def\!tkJoin#1#2{%
  \!taDataColumnTemplate{#2#1}}%

\NewFormatKey a#1{%
  \!taDataColumnTemplate\!thx{\the\!taDataColumnTemplate #1}%
  \ReadFormatKeys}

\NewFormatKey \{{%
  \!taDataColumnTemplate=\!thx{\!thx{\the\!taDataColumnTemplate}}%
  \ReadFormatKeys}

\NewFormatKey *#1#2{%
  \!taCountA=#1\relax
  \!taToksA={}%
  \!thLoop
    \ifnum \!taCountA > 0
    \!taToksA\!thx{\the\!taToksA #2}%
    \advance\!taCountA -1
  \repeat
  \!thx\ReadFormatKeys\the\!taToksA}

\NewFormatKey \LeftGlue#1{%
  \!taLeftGlue{#1}%
  \ReadFormatKeys}

\NewFormatKey \RightGlue#1{%
  \!taRightGlue{#1}%
  \ReadFormatKeys}

\NewFormatKey c{%
  \ReadFormatKeys
  \LeftGlue\hfil
  \RightGlue\hfil}

\NewFormatKey l{%
  \ReadFormatKeys
  \LeftGlue{}   
  \RightGlue\hfil}

\NewFormatKey r{%
  \ReadFormatKeys
  \LeftGlue\hfil
  \RightGlue{}}

\NewFormatKey k{%
  \h@true
  \v@true
  \!tgGetValue{\!tgInsertKern}}

\NewFormatKey i{%
  \h@true
  \v@false
  \!tgGetValue{\!tgInsertKern}}

\NewFormatKey j{%
  \h@false
  \v@true
  \!tgGetValue{\!tgInsertKern}}

\NewFormatKey n{%
  \def\!tnStyle{}%
   \futurelet\!tnext\!tnTestForBracket}

\NewFormatKey N{%
  \def\!tnStyle{$}%
   \futurelet\!tnext\!tnTestForBracket}

\NewFormatKey m{%
  \ReadFormatKeys b$ a$}

\NewFormatKey M{%
  \ReadFormatKeys \{ b{$\displaystyle} a$}

\NewFormatKey \m{%
  \ReadFormatKeys l b{{}} m}

\NewFormatKey \M{%
  \ReadFormatKeys l b{{}} M}

\NewFormatKey f#1{%
  \ReadFormatKeys b{#1}}

\NewFormatKey B{%
  \ReadFormatKeys f\bf}

\NewFormatKey I{%
  \ReadFormatKeys f\it}

\NewFormatKey S{%
  \ReadFormatKeys f\sl}

\NewFormatKey R{%
  \ReadFormatKeys f\rm}

\NewFormatKey T{%
  \ReadFormatKeys f\tt}

\NewFormatKey p{%
  \!tgGetValue{\!tgSetUpParBox}}

\NewFormatKey w{%
  \!tkTestForBeginFormat w{\!tgGetValue{\!tfSetWidth}}}

\NewFormatKey s{%
  \!taOnceOnlyTabskipfalse    
  \!tkTestForBeginFormat t{\!tgGetValue{\!tfSetTabskip}}}

\NewFormatKey o{%
  \!taOnceOnlyTabskiptrue
  \!tkTestForBeginFormat o{\!tgGetValue{\!tfSetTabskip}}}

\NewFormatKey |{%
  \!tkTestForBeginFormat |{\!tgGetValue{\!tfSetVrule}}}

\NewFormatKey \|{%
  \!tkTestForBeginFormat \|{\!tfSetAlternateVrule}}

\NewFormatKey .{%
  \!tkTestForBeginFormat.{\!tfFinishFormat}}

\NewFormatKey \EndFormat{%
  \!tkTestForBeginFormat\EndFormat{\!tfFinishFormat}}

\NewFormatKey ]{%
  \!tkTestForReFormat ] \!tfEndReFormat}

\def\!tkTestForBeginFormat#1#2{%
  \if!taBeginFormat
    \def\!ttemp{#2}%
    \!thx \!ttemp
  \else
    \toks0={#1}%
    \toks2=\!thx{\string\ReFormat}%
    \!thx \!tkImproperUse
  \fi}

\def\!tkTestForReFormat#1#2{%
  \if!taBeginFormat
    \toks0={#1}%
    \toks2=\!thx{\string\BeginFormat}%
    \!thx \!tkImproperUse
  \else
    \def\!ttemp{#2}%
    \!thx \!ttemp
  \fi}

\def\!tkImproperUse{%
  \!thError{\!thReadErrorMsg\!tkBadUseA "\the\toks0 "}%
    {\!thReadErrorMsg\!tkBadUseB \the\toks2 \space command.
    ^^J\!thReadErrorMsg\!tkBadKey}%
  \!tkReplaceKeyA}

\!thStoreErrorMsg\!tkBadUseA{Improper use of key }
\!thStoreErrorMsg\!tkBadUseB{%
  The key mentioned above can't be used in a }

\def\!tnTestForBracket{%
  \ifx [\!tnext
    \!thx\!tnGetArgument
  \else
    \!thx\!tnGetCode
  \fi}

\def\!tnGetCode#1 {
  \!tnConvertCode #1..!}

\def\!tnConvertCode #1.#2.#3!{%
  \begingroup
    \aftergroup\edef \aftergroup\!ttemp \aftergroup{%
      \aftergroup[%
      \!taCountA #1
      \!thLoop
        \ifnum \!taCountA>0
        \advance\!taCountA -1
        \aftergroup0
      \repeat
      \def\!ttemp{#3}%
      \ifx\!ttemp \empty
      \else
        \aftergroup.
        \!taCountA #2
        \!thLoop
          \ifnum \!taCountA>0
          \advance\!taCountA -1
          \aftergroup0
        \repeat
      \fi
      \aftergroup]\aftergroup}%
    \endgroup\relax
    \!thx\!tnGetArgument\!ttemp}

\def\!tnGetArgument[#1]{%
  \!tnMakeNumericTemplate\!tnStyle#1..!}

\def\!tnMakeNumericTemplate#1#2.#3.#4!{
  \def\!ttemp{#4}%
  \ifx\!ttemp\empty
    \!taDimenC=0pt
  \else
    \setbox0=\hbox{\m@th #1.#3#1}%
    \!taDimenC=\wd0
  \fi
  \setbox0 =\hbox{\m@th #1#2#1}%
  \!thToksEdef\!taDataColumnTemplate={%
    \noexpand\!tnSetNumericItem
    {\the\wd0 }%
    {\the\!taDimenC}%
    {#1}%
    \the\!taDataColumnTemplate}  
  \ReadFormatKeys}

\def\!tnSetNumericItem #1#2#3#4 {
  \!tnSetNumericItemA {#1}{#2}{#3}#4..!}

\def\!tnSetNumericItemA #1#2#3#4.#5.#6!{%
  \def\!ttemp{#6}%
  \hbox to #1{\hss \m@th #3#4#3}%
  \hbox to #2{%
    \ifx\!ttemp\empty
    \else
       \m@th #3.#5#3%
    \fi
    \hss}}

\def\MakeStrut#1#2{%
  \vrule width0pt height #1 depth #2}

\def\StandardTableStrut{%
  \MakeStrut{\StrutHeightFactor\StrutUnit}
    {\StrutDepthFactor\StrutUnit}}

\def\AugmentedTableStrut#1#2{%
  \dimen@=\StrutHeightFactor\StrutUnit
  \advance\dimen@ #1\StrutUnit
  \dimen@ii=\StrutDepthFactor\StrutUnit
  \advance\dimen@ii #2\StrutUnit
  \MakeStrut{\dimen@}{\dimen@ii}}

\def\Enlarge#1#2{
  \!taDimenA=#1\relax
  \!taDimenB=#2\relax
  \let\!TsSpaceFactor=\empty
  \ifmmode
    \!thx \mathpalette
    \!thx \!TsEnlargeMath
  \else
    \!thx \!TsEnlargeOther
  \fi}

\def\!TsEnlargeOther#1{%
  \ifhmode
    \setbox\z@=\hbox{#1%
      \xdef\!TsSpaceFactor{\spacefactor=\the\spacefactor}}%
  \else
    \setbox\z@=\hbox{#1}%
  \fi
  \!TsFinishEnlarge}

\def\!TsEnlargeMath#1#2{%
  \setbox\z@=\hbox{$\m@th#1{#2}$}%
  \!TsFinishEnlarge}

\def\!TsFinishEnlarge{%
  \dimen@=\ht\z@
  \advance \dimen@ \!taDimenA
  \ht\z@=\dimen@
  \dimen@=\dp\z@
  \advance \dimen@ \!taDimenB
  \dp\z@=\dimen@
  \box\z@ \!TsSpaceFactor{}}

\def\OpenUp#1#2{%
  \advance \StrutHeightFactor #1\relax
  \advance \StrutDepthFactor #2\relax}

\def\BeginTable{%
  \futurelet\!tnext\!ttBeginTable}

\def\!ttBeginTable{%
  \ifx [\!tnext
    \def\!tnext{\!ttBeginTableA}%
  \else
    \def\!tnext{\!ttBeginTableA[c]}%
  \fi
  \!tnext}

\def\!ttBeginTableA[#1]{%
  \if #1u
    \ifmmode
      \def\!ttEndTable{
        \relax}
    \else                   
      \bgroup
      \def\!ttEndTable{%
        \egroup}%
    \fi
  \else
    \hbox\bgroup $
    \def\!ttEndTable{%
      \egroup 
      $
      \egroup}
    \if #1t%
      \vtop
    \else
      \if #1b%
        \vbox
      \else
        \vcenter 
      \fi
    \fi
    \bgroup      
  \fi
  \advance\!taRecursionLevel 1 
  \let\!ttRightGlue=\relax  
  \everycr={}
  \ifnum \!taRecursionLevel=1
    \!ttInitializeTable
  \fi}

\bgroup
  \catcode`\|=13
  \catcode`\"=13
  \catcode`\~=13
  \gdef\!ttInitializeTable{%
    \let\!ttTie=~ 
    \let\!ttDH=\- 
    \catcode`\|=\active
    \catcode`\"=\active
    \catcode`\~=\active
    \def |{\unskip\!ttRightGlue&&}
    \def\|{\unskip\!ttRightGlue&\omit\!ttAlternateVrule}%
    \def"{\unskip\!ttRightGlue&\omit&}
    \def~{\kern .5em}
    \def\\{\!ttEndOfRow}%
    \def\-{\!ttShortHrule}%
    \def\={\!ttLongHrule}%
    \def\_{\!ttFullHrule}%
    \def\Left##1{##1\hfill\null}
    \def\Center##1{\hfill ##1\hfill\null}
    \def\Right##1{\hfill##1}%
    \def\use{\!ttuse}%
    \the\EveryTable}
\egroup

\let\!ttRightGlue=\relax  

\def\!ttDoHalign{%
  \baselineskip=0pt \lineskiplimit=0pt \lineskip=0pt %
  \tabskip=0pt
  \halign \the\!taTableSpread \bgroup
   \span\the\!taPreamble
   \ifx \!tfRowOfWidths \empty
   \else
     \!tfRowOfWidths \cr %
   \fi}

\def\EndTable{%
  \egroup 
  \!ttEndTable}

\def\!ttEndOfRow{%
  \futurelet\!tnext\!ttTestForBlank}

\def\!ttTestForBlank{%
  \ifx \!tnext\!thSpaceToken  
    \!thx\!ttDoStandard
  \else
    \!thx\!ttTestForZero
  \fi}

\def\!ttTestForZero{%
  \ifx 0\!tnext
    \!thx \!ttDoZero
  \else
    \!thx \!ttTestForPlus
  \fi}

\def\!ttTestForPlus{%
  \ifx +\!tnext
    \!thx \!ttDoPlus
  \else
    \!thx \!ttDoStandard
  \fi}

\def\!ttDoZero#1{
  \cr}

\def\!ttDoPlus#1#2#3{
  \AugmentedTableStrut{#2}{#3}%
  \cr}

\def\!ttDoStandard{%
  \StandardTableStrut
  \cr}

\def\!ttAlternateVrule{%
  \!tgGetValue{\!ttAVTestForCode}}  

\def\!ttAVTestForCode{%
  \ifnum \!tgCode=2              
    \!thx\!ttInsertVrule         
  \else
    \!thx\!ttAVTestForEmpty
  \fi}

\def\!ttAVTestForEmpty{%
  \ifx \!tgValue\empty           
    \!thx\!ttAVTestForBlank
  \else
    \!thx\!ttInsertVrule         
  \fi}

\def\!ttAVTestForBlank{%
  \ifx \!ttemp\!thSpaceToken     
    \!thx\!ttInsertVrule
  \else
    \!thx\!ttAVTestForStar
  \fi}

\def\!ttAVTestForStar{%
  \ifx *\!ttemp                  
    \!thx\!ttInsertDefaultPR     
  \else
    \!thx\!ttGetPseudoVrule       
  \fi}

\def\!ttInsertVrule{%
  \hfil
  \vrule \!thWidth
    \ifnum \!tgCode=1
      \ifx \!tgValue\empty
        \LineThicknessFactor
      \else
        \!tgValue
      \fi
      \LineThicknessUnit
    \else
      \!tgValue
    \fi
  \hfil
  &}

\def\!ttInsertDefaultPR*{%
  \PseudoVrule    
  &}

\def\!ttGetPseudoVrule#1{%
  \toks0={#1}%
  #1&}

\def\PseudoVrule{}

\def\!ttuse#1{%
  \ifnum #1>\@ne
    \omit
    \mscount=#1 
    \advance\mscount by \m@ne
    \advance\mscount by \mscount
    \!thLoop
      \ifnum\mscount>\@ne
      \sp@n 
    \repeat
    \span
  \fi}

\def\!ttUse#1[{%
  \!ttuse{#1}%
  \ReFormat[}

\def\!ttFullHrule{%
  \noalign
  \bgroup
  \!tgGetValue{\!ttFullHruleA}}

\def\!ttFullHruleA{%
  \!ttGetHalfRuleThickness 
  \hrule \!thHeight \dimen0 \!thDepth \dimen0
  \penalty0 
  \egroup} 

\def\!ttShortHrule{%
  \omit
  \!tgGetValue{\!ttShortHruleA}}

\def\!ttShortHruleA{%
  \!ttGetHalfRuleThickness 
  \leaders \hrule \!thHeight \dimen0 \!thDepth \dimen0 \hfill
  \null    
  \ignorespaces}

\def\!ttLongHrule{%
  \omit\span\omit\span \!ttShortHrule}

\def\!ttGetHalfRuleThickness{%
  \dimen0 =
    \ifnum \!tgCode=1
      \ifx \!tgValue\empty
        \LineThicknessFactor
      \else
        \!tgValue    
      \fi
      \LineThicknessUnit
    \else
      \!tgValue      
    \fi
  \divide\dimen0 2 }

\def\WidenTableBy#1{%
  \ifdim #1=0pt
    \!taTableSpread={}%
  \else
    \!taTableSpread={spread #1}%
  \fi}

\def\JustLeft{%
  \omit \let\!ttRightGlue=\hfill}
\def\JustCenter{%
  \omit \hfill\null \let\!ttRightGlue=\hfill}

\let\\=\!tacr
\catcode`\!=12
\catcode`\@=12

\documentstyle[12pt,fleqn,lists,epsfig]{article} 
\def\ie{{\it i.e.}}  

\def\stot{\mbox{$\sigma_{\rm tot}$}}         
\def\pbar{\mbox{$\bar {\rm p}$}}

\def\C{\JustCenter} 
\def\LC#1{\C \Lower{#1}} 
\def\PseudoVrule{\hfil \vrule \hskip2pt \vrule \hfil} 
\begin{document}    
%
%
%
\newcommand\reportnumber{389} 
\newcommand\mydate{October, 1994} 
\newlength{\nulogo} 
\settowidth{\nulogo}{\small\sf{N.U.H.E.P. Report No. \reportnumber}}
\title{
\vspace{-.8in} 
\hfill\fbox{{\parbox{\nulogo}{\small\sf{Northwestern University: \\
N.U.H.E.P. Report No. \reportnumber\\
          \mydate}}}}
          \vspace{.5in} \\
{The High Energy Behavior of the Forward Scattering Parameters---\\
 \stot, $\rho$, and $B$}
}
\author{
M.~M.~Block
\thanks{to be published in the Proceedings of the XXIV International Symposium
on Multiparticle
Dynamics, Eds. A.~Giovannini, S.~Lupia and R.~Ugocionni, World Scientific,
Singapore.
Work partially supported by Department of Energy grant
DOE 0680-300-N008 Task B.}\vspace{-5pt}   \\
{\small\em Department of Physics and Astronomy,} \vspace{-5pt} \\ 
{\small\em Northwestern University, Evanston, IL 60208}\\
\vspace{-5pt}\\
%
F.~Halzen
\thanks{Work partially supported by
Department of Energy contract
DE-AC02-76ER0088 and
the University of Wisconsin Research
Committee with funds granted by the Wisconsin Alumni Research Foundation.
}\vspace{-5pt} \\
{\small\em Department of Physics,} \vspace{-5pt} \\
{\small\em University of Wisconsin, Madison, WI 53706}  \\
\vspace{-5pt}\\
%
B.~Margolis
\thanks{Work partially supported by
the Natural Sciences and Engineering Research Council of Canada and the FCAR
of the Province of Quebec.
}\vspace{-5pt} \\
{\small\em  Physics Department,}\vspace{-5pt}  \\
{\small\em McGill University, Montreal, Canada H3A 2T8}\\
\vspace{-5pt}\\
%
A.~R.~White
\thanks{Work supported by
Department of Energy contract
W-31-109-ENG-38.}\vspace{-5pt} \\
{\small\em High Energy Physics Division,} \vspace{-5pt} \\
{\small\em Argonne National Laboratory, Argonne, Il 60439}  \\
}    
\date{} 
\maketitle
\date{} 
\vspace{-.3in}

\renewcommand\thepage{\ }
%
\begin{abstract} 
Utilizing the most recent experimental data, we reanalyze high energy \pbar p
and pp data, using two
distinct (and {\em dissimilar}) analysis techniques: (1) asymptotic amplitude
analysis, under the
assumption that we have reached `asymptopia', and (2) an eikonal model
whose amplitudes are designed to mimic real QCD amplitudes.  The former gives
strong
evidence for a $\log \,(s/s_0)$ dependence at {\em current} energies
and {\em not} $\log^2 (s/s_0)$, and
demonstrates that odderons are {\em not} necessary to explain the
experimental data.  The latter gives a unitary model for extrapolation into
true `asymptopia' from current energies, allowing us to
predict the values of the total cross section at future supercolliders.
Using our QCD-model, we obtain $\stot(16\,\, {\rm  TeV})=109\pm4$\,mb
and $\stot(40\,\, {\rm TeV})=124\pm4$\,mb.

\end{abstract}  
%
\pagenumbering{arabic}
\renewcommand{\thepage}{-- \arabic{page}\ --}  
\renewcommand{\thesection}{\Roman{section}}  

\section{Introduction}
Recently, the CDF group has announced new
cross sections at $\sqrt s=546$ GeV and 1080 GeV.  The
result at 1800 GeV is at variance with their earlier value (announced
at the `Blois'
meeting at Elba).  Only the new results of CDF have been included in
Table~{\ref{ta:exptresults}, along with
the recent precision remeasurement of $\rho$ at $\sqrt s=546$ GeV
by UA4/2 at the S\pbar pS CERN Collider, the reanalysis of the
$\sqrt s=1800$ GeV Tevatron Collider data by E710 of
$\rho$, $\stot$ and
$B$, as well as the new E710 measurement at the Tevatron
of $\stot$ and
$B$ at $\sqrt s=1020$ GeV.  These results complete the information we will
have on high energy forward  scattering parameters until the LHC is
operative.  Although the CDF and E710 results at 1800 GeV disagree, we see
that the inclusion of both of them does {\em not} change the conclusions
found in our earlier analysis\cite{aspensig}, given at Multiparticle/93 in
Aspen.
\begin{table}[htbp]  
%
\caption{{\protect\small Recent Experimental Results
at High Energies}
\label{ta:exptresults}} 
$$          
\BeginTable
     \def\Energy{$\sqrt s $}
     \BeginFormat
 |  ck    |         ck     |   ck          | ck         |   ck     |
     \EndFormat
 "{}& \use4 \= &\\0
 "{}    |\C{$\sigma$}   | \C{$B$}          |\LC{$\rho$}   |  \C{\Energy} |
\\+20
 " {}   | (mb)          |\C{$\mbox{(GeV/c)}^{-2}$} |{}    |\C(GeV) |\\+02
 & \use5 \= &\\0
 |E710 | $72.2\pm 2.7$ |$16.72\pm 0.44$ |$0.134\pm 0.069$
          |1800|\\+40
 &\use5 \=&\\0
 |E710 | $61.6\pm 5.7$ |$16.2\pm 0.70$ |{}
          |1020|\\+40
&\use5 \=&\\0
 |CDF | $80.0\pm 2.2$ |$16.98\pm 0.25$ |{}
          |1800|\\+40
&\use5 \=&\\0
 |CDF | $61.9\pm 1.5$ |$15.28\pm 0.58$ |{}
          |546|\\+40
 &\use5 \=&\\0
 |UA4/2 | {} |  {} | ${0.135\pm 0.02} $         |546|
 \\+20
& \use5 \= & \\0
\EndTable
$$
\end{table}
We present two approaches to interpret the high energy data:
\begin{romList}
\item Model 1---An analytic asymptotic amplitude
analysis,
\item Model 2---A QCD-inspired model.
\end{romList}

\section{Asymptotic Amplitude Analysis}
In spite of the fact that there are excellent arguments
\cite{others}
that the
energy region in which present experiments are conducted---even at the
Tevatron Collider---is too low to be
considered asymptotic, we will consider here the consequences of assuming
the {\em opposite}. This allows us to test specific hypotheses
using a well-defined phenomenological
analysis .
We caution the reader that we
{\em don't believe} we are in `asymptopia' and thus {\em don't believe} the
analysis is applicable as a true asymptotic analysis. We {\em do believe}
that present day energies are too low to make  a truly asymptotic analysis.
Nonetheless, we feel that such an analysis is valuable as a guideline to
what is and is not happening at present energies.

We apply a ``standard'' asymptotic analytic amplitude
analysis 
pro\-ce\-dure\cite{others} to the now-available data
on $\stot$, the total cross section and $\rho$, the ratio of the real
to the imaginary portion of the forward scattering amplitude, in the energy
region $\sqrt s$ = 5 to 1800
GeV. The data are parameterized in terms of even and odd analytic
amplitudes. Consistent with all asymptotic theorems, this allows use of
even amplitudes varying as fast as $\log^{2}(s/s_0)$ and odd
amplitudes (the `Odderon' family) that do {\em not} vanish as
$s\rightarrow\infty$.

We show here only the large $s$ limit of the even and odd
amplitudes that are used\cite{others}. We make five fits to the data:
\begin{romList}
\item Fit 1: $\log^{2}(s/s_0)$ energy dependence for the cross section, with
no  Odderon amplitude,
\item Fit 2: $\log^{2}(s/s_0)$ energy dependence for the cross section, with
an Odderon amplitude whose cross sectional dependence is $\log s$, the most
rapid behavior allowed by asymptotic theorems,
\item Fit 3: $\log^{2}(s/s_0)$ energy dependence for the cross section, with
an Odderon amplitude whose cross sectional dependence is constant,
\item Fit 4: $\log \,(s/s_0)$ energy dependence for the cross section, with
no  Odderon amplitude,
\item Fit 5: $\log \,(s/s_0)$ energy dependence for the cross section, with
an Odderon amplitude whose cross sectional dependence  is constant, the most
rapid behavior allowed by asymptotic theorems for this choice of even
amplitude.
\end{romList}

In all cases, an odd amplitude which vanishes with increasing energy is
also employed, as well as an even amplitude that mimics Regge behavior.
\subsection{log$\,{}^{2}(s)$ Energy Behavior}
We introduce $f_+$ and $f_-$, the even and odd (under crossing) analytic
amplitudes
at $t=0$, and define the $\bar {\rm p}$p and pp forward scattering
amplitudes by
$f_{\bar{\rm p}{\rm p}}= f_{+} + f_{-}\,\,\,\,{\rm and}\,\,\,\,
f_{\rm pp}=f_+ - f_-,$
giving total cross sections $\stot$ and
the $\rho $-values
\begin{eqnarray}
\sigma_{\bar{\rm p}{\rm p}}=\frac{4\pi}{p}
                    \,{\rm Im}\,f_{\bar{\rm p}{\rm p}}
,\,\,\,\,
\sigma_{{\rm pp}}=\frac{4\pi}{p}\,{\rm Im}\,f_{\rm pp},\,\,\,\,
\rho_{\bar{\rm p}{\rm p}}=
\frac{ {\rm Re}\,f_{\bar{\rm p}{\rm p}} }
{ {\rm Im}\,f_{ \bar{\rm p}{\rm p} } }
,\,\,\,\, {\rm and}\,\,\,\,
\rho_{{\rm pp}}=
\frac{ {\rm Re}\,f_{{\rm pp}} }
{ {\rm Im}\,f_{{\rm pp}} }.
\end{eqnarray}
We parameterize the `conventional' even and odd amplitudes $f_+$ and $f_-$
by :
\begin{eqnarray}
     \frac{4\pi}{p}f_+ &=& i \left ( A+
     \beta \left [\log \left (\frac{s}{s_0}\right ) - i \frac{\pi}{2}
     \right ]^2 + c\,s^{\mu -1}e^{i\pi(1-\mu )/2}\right ) \label{eq:even2} \\
     \frac{4\pi}{p}f_- &=& -Ds^{\alpha -1} e^{i\pi (1-\alpha )/2}.
\label{eq:odd}
\end{eqnarray}

The parameter $\alpha $ in Eq~(\ref{eq:odd}) turns out to be about 0.5,
and thus
this odd amplitude vanishes as $s\rightarrow\infty$.

Asymptotic theorems by Eden and Kinoshita\cite{others} prove
that the {\em difference} of cross sections can not grow
faster than $\log^{\gamma /2}(s)$, when the cross section grows as
$\log^{\gamma }(s)$. Thus, odd amplitudes which do {\em not}
vanish as
$s\rightarrow\infty$ for this case are :

     $\frac{4\pi}{p}f_-^{(0)}= -\,\epsilon ^{(0)},\,\,\,\,
     \frac{4\pi}{p}f_-^{(1)} =-\left [
                      \log \left ( \frac {s}{s_0} \right ) -i \frac{\pi}{2}
                       \right ] \epsilon ^{(1)},\,\,\,\, {\rm and},\,\,\,\,
     \frac{4\pi}{p}f_-^{(2)} = -\left [
                      \log \left ( \frac {s}{s_0} \right ) -i \frac{\pi}{2}
                      \right ]^{2} \epsilon ^{(2)}.$

The complete odd amplitude is formed by adding any one (or none)
of the $f^{(i)}_-$  to the conventional odd amplitude $f_-$
of Eq~(\ref{eq:odd}).  We then fit the experimental $\rho $ and
$\stot $ data, for
both pp and $\bar{\rm p}$p, for energies between 5 and 1800 GeV, to obtain
the real constants $A,\beta ,s_0,c,\mu ,D,\alpha ,\epsilon ^{(i)}$.  The data
used below 500 GeV are listed in \cite{others}, and the high energy
points are from UA1, UA4, E710
and CDF\cite{others}.
We emphasize that what we really fit for the UA4 and CDF cross sections is
the measured
experimental quantity $\stot\times (1+\rho^2)$, which is appropriate for
experiments that measure a `luminosity-free' cross section, whereas for UA1
and the 1020 GeV point of E710,
we fit the experimental quantity $\stot\times \sqrt {1+\rho^2}$,
which was
their experimentally measured quantity (they measured a
`luminosity-dependent' cross section).

\subsection{Fitted Results for log$\,{}^{2}(s)$ Behavior}
\begin{romList}
\item Fit 1---This fit uses no Odderons in the odd amplitude and uses the
even amplitude of Eq~(\ref{eq:even2}).
The $\chi^2$/d.f. ($\chi^2$/degree of freedom)
for the fit
was 2.03, a rather large number.  The fitted constants are shown in
Table \ref{ta:amp},
Fit 1---the computed
curves are shown in Fig.~\ref{sfit1}(for $\stot$) and Fig.~\ref{rfit1} (for
$\rho$).
\noindent The most obvious features of the fit are
\begin{alphList}
 \item the predicted value
of the total cross section is much
too high
to fit the experimental values (E710 and CDF) at 1800 GeV,
\item it predicts much too high a $\rho $-value at 546 GeV.
\end{alphList}

We conclude that a simple $\log^2(s)$ fit does not fit the data.

\item Fit 2---
We fit the data with an additional degree of
freedom, by adding Odderon 2  to $f_-$ of
Eq~(\ref{eq:odd}), along with the even
amplitude of Eq~(\ref{eq:even2}).  The parameters are summarized as Fit 2,
in Table \ref{ta:amp}. Again, we
conclude that this combination doesn't fit the data, since the high energy
cross section predicted at 1800 GeV is much too high.
Although the
$\rho$-value predicted at 540 GeV is slightly lower, the $\rho$ values
predicted are still too high.
\item Fit 3---The odd
amplitude added to the conventional $f_-$ of Eq~(\ref{eq:odd}) was
Odderon 1.  The parameters are given as Fit 3 in
Table \ref{ta:amp}.  Again, the fit suffers
from the
same defect as the Odderon 2 fit, giving much too high a total cross section
at 1800 GeV, as well as predicting a UA4/2
$\rho$-value which was much too high.

The addition of Odderon 0 can have no effect on
the cross section. Since it turns out to have a negligible effect on $\rho$,
we will not
consider it further.
\end{romList}
We conclude that an even amplitude varying as
$\log^2(s/s_0)$ {\em does not fit} the cross section data. We see that
the experimental cross section does not rise as rapidly as $\log^2(s/s_0)$,
in the present-day energy region. The addition
of an Odderon term does not change this conclusion.\newpage
\begin{figure}[ht]
\centerline{\psfig{figure=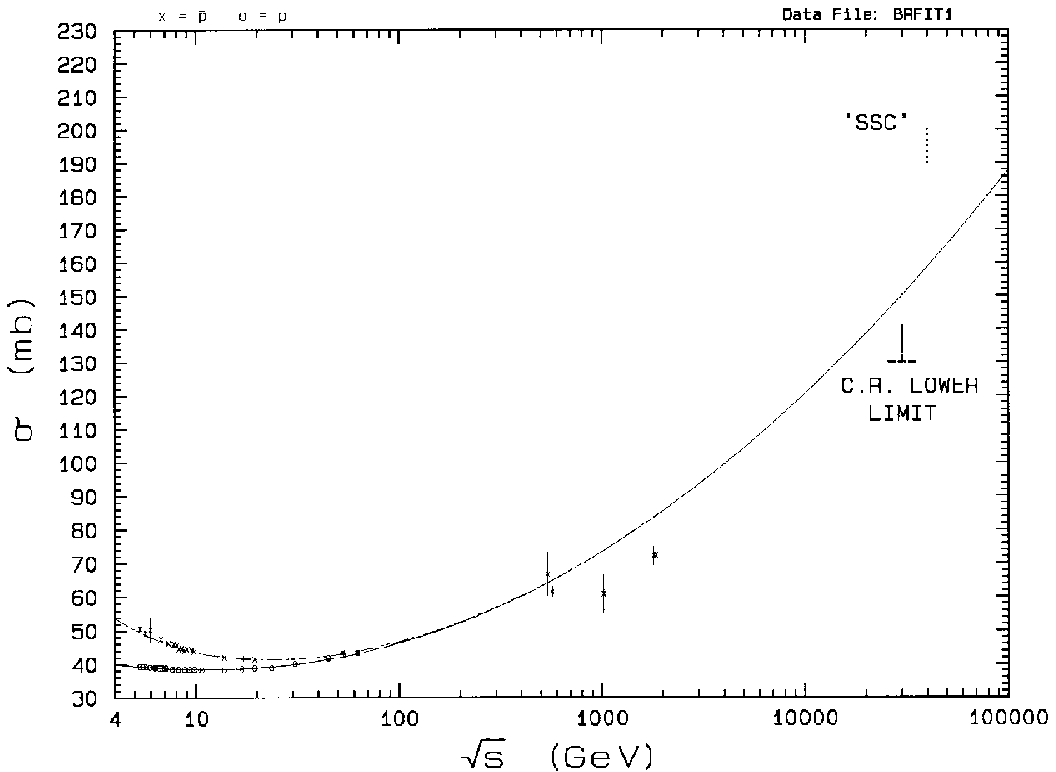,width=4.25in}}
\caption{The total cross section  $\sigma_{tot}$, in mb, for
$\bar {\rm{p}}$p and
pp scattering {\it vs.} the energy, $\protect\sqrt s$, in GeV, for Fit 1,
described
in Table I.  The fit was made with a $\log^2(s)$ energy variation, and no
Odderon.  The crosses are for the $\bar {\rm{p}}$p experimental data and
the circles
indicate pp data. The dot-dashed curves are for $\bar {\rm{p}}$p, and the
solid curves for pp. The pp cosmic-ray lower limit\protect\cite{others} is
appended to
the curve, but is {\em not} used in the fit.\protect\label{sfit4}}
\end{figure}
\begin{figure}[hb]
\centerline{\psfig{figure=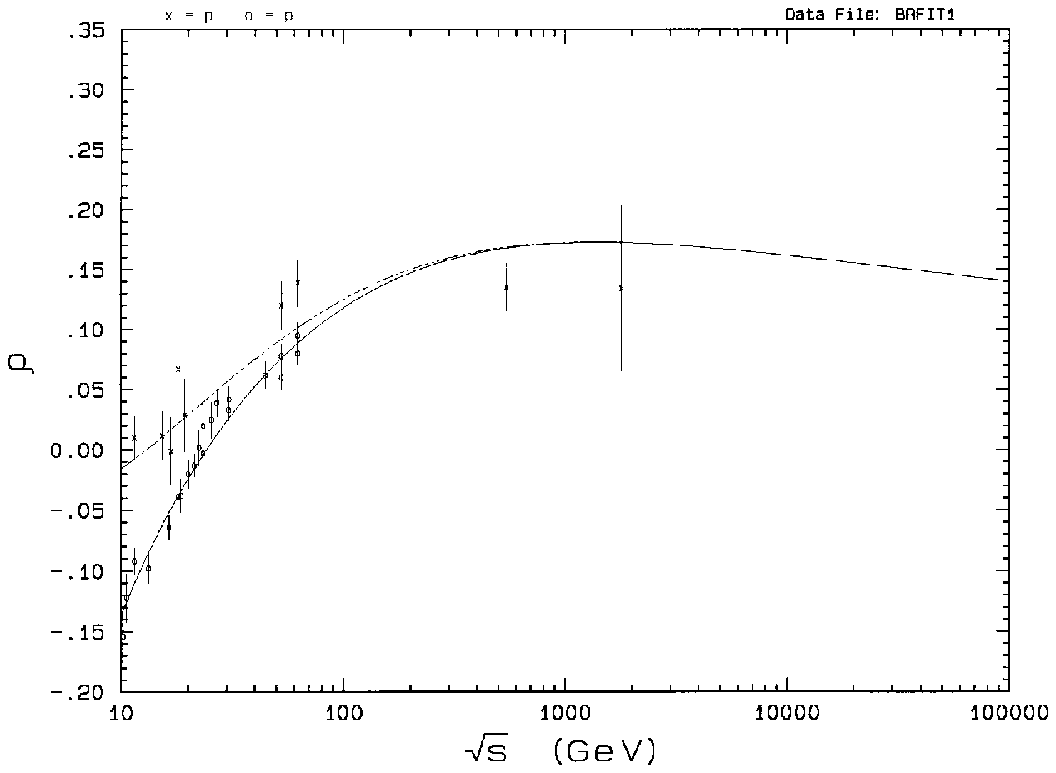,width=4.25in}}
\caption{The $\rho$-value for $\bar {\rm{p}}$p and
pp scattering {\it vs.} the energy, $\protect\sqrt s$, in GeV, for Fit 1,
described
in Table I.  The fit was made with a $\log^2(s)$ energy variation, and no
Odderon. The crosses are for the $\bar {\rm{p}}$p experimental data and
the circles
indicate pp data. The dot-dashed curves are for $\bar {\rm{p}}$p, and the
solid curves for pp.\protect\label{rfit4}
}
\end{figure}

\begin{table}[h,t]                   
%
\def\arraystretch{1.5}            
\begin{tabular}[b]{|l||l|l|l||l|l|}
     \cline{2-6}
      \multicolumn{1}{c|}{}
      &\multicolumn{3}{c||}{$\stot \sim \log^2(s/s_0)$}
      &\multicolumn{2}{c|}{$\stot \sim \log(s/s_0)$}\\
      \hline
      Parameters&Fit 1&Fit 2&Fit 3&Fit 4&Fit 5 \\ \hline
     $A$ (mb)&$40.3\pm .21$&$40.1\pm$ .25&$41.6\pm$ .04&$-.6\pm 1.4$
     &$-16.9\pm 8.8$\\
     $\beta$ (mb)&$.47\pm .02$&$.47\pm .02$&$.57\pm .01$&$7.7\pm .1$
     &$9.0\pm .7$ \\
     $s_0$ (${\rm (GeV)}^2$)&$201\pm 21$&$198\pm 24$&$345\pm 11$&$500$
     &$500$ \\
     $D$ (mb${\rm (GeV)}^{2(1-\alpha)}$)&$-40.6\pm 1.9$&$-39.1\pm 2.1$
     &$-37.2\pm 1.7$&$-44.2\pm 2.1$&$-43.3\pm2.1$ \\
     $\alpha$&$.46\pm .02$&$.47\pm .02$&$.48\pm .02$&$.44\pm .01$
     &$.45\pm .02$ \\
     $c$ (mb${\rm (GeV)}^{2(1-\mu)})$&$30.4\pm 4.3$&$28.0\pm 4.2$
     &$6.6\pm 2.9$&$124\pm 2$&$146\pm 12$ \\
     $\mu$&$.46$&$.49$&$.49$&$.83\pm .01$&$.85\pm .01$ \\
     $\epsilon^{(2)}$ (mb)&&$-.022 \pm.011$&&& \\
     $\epsilon^{(1)}$ (mb)&&&$.040 \pm.040$&&$-.013\pm .042$ \\
     \hline
     $\chi^2/$d.f.&2.03&1.93&2.66&1.26&1.24 \\
     d.f.&82&81&81&82&81\\
     \hline
\end{tabular}
     \caption{\protect\small Results of fits to total cross sections and
     $\rho $-values, including Odderons. Fit 1, Fit 2 and Fit 3 correspond
     to an asymptotic
     cross section variation of $\log ^2(s/s_0)$, with no Odderon,
     Odderon~2 and
     Odderon~1, respectively, whereas Fit 4 and Fit 5 correspond
     to an energy dependence of $\log (s/s_0)$, with no Odderon and Odderon~1,
     respectively.\label{ta:amp}}
\end{table}
\def\arraystretch{1}  

\subsection{log$\,(s)$ Energy Behavior}
Since the experimental cross section in the energy region 5-1800 GeV did
not vary as fast as $\log ^2(s/s_0)$, we now consider an asymptotic variation
that goes as $\log \,(s/s_0)$.  We substitute for the even
amplitude in
Eq~(\ref{eq:even2}) a new amplitude $f_+$ varying as $\log \,(s/s_0)$,
$
     \frac{4\pi}{p}f_+ = i \left (A+
     \beta\left [\log \left (\frac{s}{s_0}\right ) - i \frac{\pi}{2}
     \right ] + c\,s^{\mu -1}e^{i\pi(1-\mu )/2} \right ).
$
We use the conventional odd amplitude of Eq~(\ref{eq:odd}), along with no
Odderon or Odderon 1, in Fits 4 and 5, respectively. We make the
important observation that since
the energy variation of the cross section is now only $\log (s)$,
Odderon 2 is {\em not} allowed by the asymptotic theorems.

\subsection{Fitted Results for log$\,(s)$ Behavior}
\begin{romList}
\item
Fit 4---The data are fitted with a log$\,(s/s_0)$ cross section
energy behavior,
with no Odderon. The results are detailed in Table \ref{ta:amp}, and plotted
in Fig.~\ref{sfit4} and
Fig.~\ref{rfit4}. The fit is quite satisfactory, giving a $\chi^2$/d.f.
of 1.26, fitting reasonably well to all cross section data over the
entire range of energy.  Most importantly, it now fits the UA4/2 $\rho$-value
at 546 GeV,
as well as the E710 $\rho$-value at 1800 GeV.\newpage
\begin{figure}[ht]
\centerline{\psfig{figure=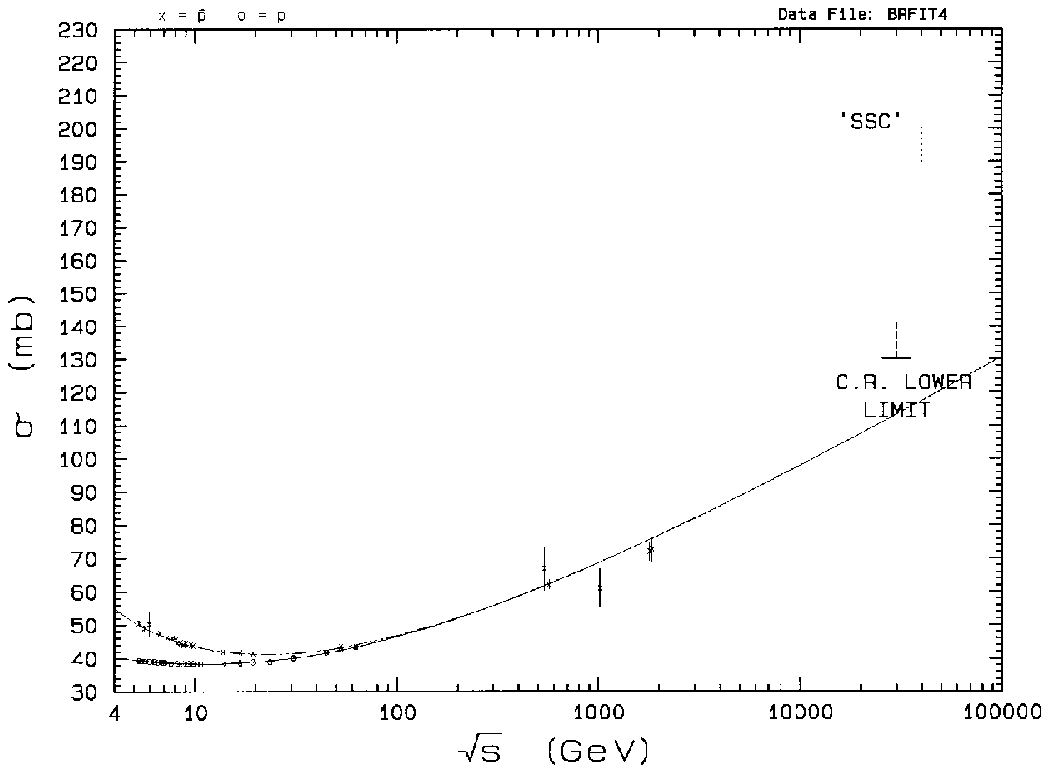,width=4.25in}}
\caption{The total cross section  $\sigma_{tot}$, in mb, for
$\bar {\rm{p}}$p and
pp scattering {\it vs.} the energy, $\protect\sqrt s$, in GeV, for Fit 4,
described
in Table I.  The fit was made with a $\log(s)$ energy variation, and no
Odderon.  The crosses are for the $\bar {\rm{p}}$p experimental data and
the circles
indicate pp data. The dot-dashed curves are for $\bar {\rm{p}}$p, and the
solid curves for pp. The pp cosmic-ray lower limit\protect\cite{others} is
appended to
the curve, but is {\em not} used in the fit.\protect\label{sfit1}}
\end{figure}
\begin{figure}[hb]
\centerline{\psfig{figure=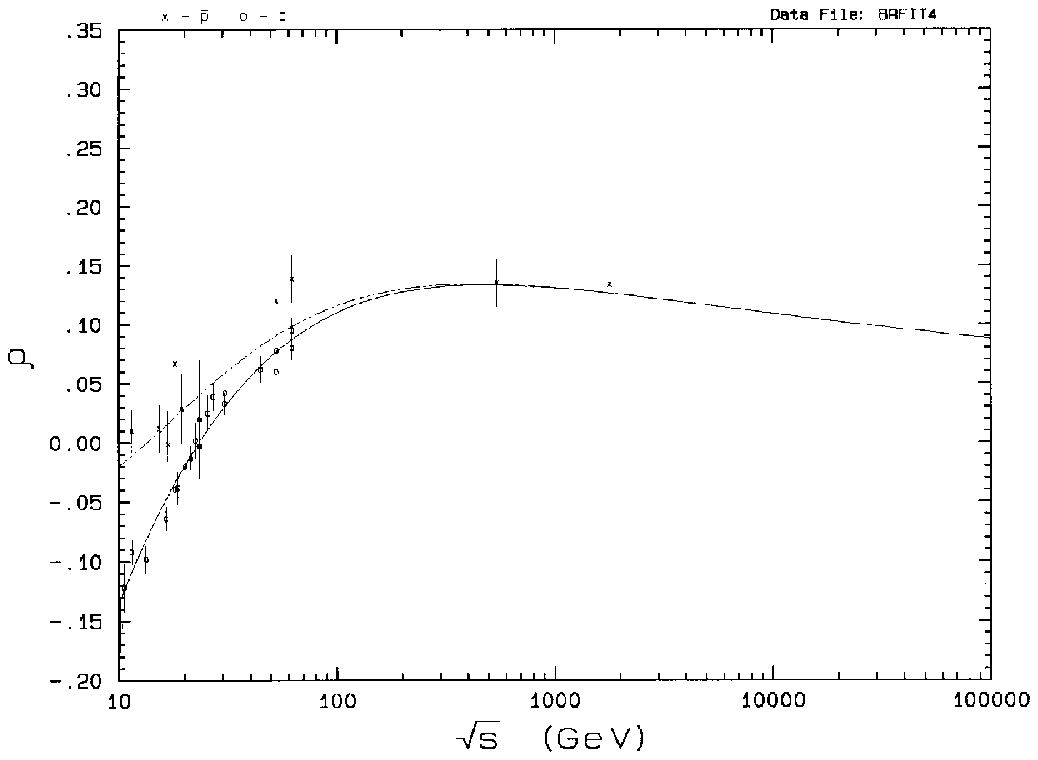,width=4.25in}}
\caption{The $\rho$-value for $\bar {\rm{p}}$p and
pp scattering {\it vs.} the energy, $\protect\sqrt s$, in GeV, for Fit 4,
described
in Table I.  The fit was made with a $\log(s)$ energy variation, and no
Odderon. The crosses are for the $\bar {\rm{p}}$p experimental data and
the circles
indicate pp data. The dot-dashed curves are for $\bar {\rm{p}}$p, and the
solid curves for pp.\protect\label{rfit1}
}
\end{figure}

Using this fit, we obtain a cross section of
$\stot=117.4\pm 1.3$\,mb at 40 TeV (where the error is
statistical, and results from the errors in the fitted parameters) and a
LHC cross section $\stot=104.4\pm 1.0$\,mb.  As we will comment on
later, we believe that we are not yet in `asymptopia', and we think that
the cross section will ultimately rise faster than $\log (s)$.  Hence,
we consider these high energy cross section extrapolations
{\em lower limits}
to the real cross sections at these high energies.
\item Fit 5---
The data are fitted with a log$\,(s)$ cross section energy behavior,
along with Odderon 1. The results are given in Table
2. This fit (as is Fit 4) is
quite satisfactory, giving a $\chi^2$/d.f. of 1.24. Indeed, it is almost
indistinguishable from fit 4.
\end{romList}
We find that the experimental cross sections and $\rho$-values in the
energy domain 5--1800 GeV can be reproduced using a
$\log (s/s_0)$ energy variation. Further, the introduction of an Odderon
amplitude is not needed to explain the experimental data.
Thus, we conclude that the Odderon
hypothesis is {\em irrelevant}, since the experimental data do
not require the introduction of non-vanishing odd amplitudes.

\section{Too Low an Energy for `Asymptopia'?}
Indications that we may still be far from asymptopia come from
experimental measurements of
the elastic differential cross section,
$\frac{d\sigma}{dt}$, out to intermediate $t$-values.
Consider the $t=0$
parameters, the slope $B$ and the curvature
parameter $C$, defined as:

$B=\left [\frac {d\,\,}{dt}\left (\log \frac{d\sigma}{dt}(t)
     \right) \right ]_{t=0} \,\,\,\,{\rm and}\,\,\,
C=\frac{1}{2} \left [\frac {d^2\,}{dt^2}\left (\log \frac{d\sigma}{dt}(t)
     \right) \right ]_{t=0} .$
Thus, we can parameterize the elastic scattering cross section for
intermediate $t$ values as:
$\frac{d\sigma_{\rm el}}{dt}(t) =\left [\frac{d\sigma_{\rm el}}{dt}(t)
\right ]_{t=0}\,e^{B|t|+Ct^2},$
for the four-momentum transfer interval $0\leq |t| \leq t_{max}$, where
$t_{max}\approx 0.5 {\rm (GeV/c)}^2$.

At low energies, at the ISR\cite{others} and the S$\bar {\rm p}$pS
\cite{others}, a substantial positive curvature $C$ was found.  In a recent
experiment, the E710 group\cite{others} measured $B=16.26\pm 0.23
{\rm (GeV/c)}^{-2}$ and $C=0.14\pm 0.70{\rm (GeV/c)}^{-4}$, indicating that
the curvature is going through zero. This suggests that the
sign of the curvature is about to change to the
{\em negative} curvature associated with a variety of asymptotic theories,
including an expanding disk \cite{others} and the Critical
Pomeron\cite{others}.
We recall to the reader  that for the particular case of a disk with
sharp edges, the
curvature (which is negative for all energies)
is given by
$C=-\frac{R^4}{192}$ and the slope by $B=\frac{R^2}{4}$, where
$R$ is the radius of the disk.

One can define the threshold to `asymptopia' as the energy where the
curvature $C$ goes through zero.
Figure 1, taken from
Block, Halzen and Margolis\cite{others},
compares their predictions for the elastic differential scattering cross
section
$\frac{d\sigma}{dt}$  {\it vs.} $|t|$
for $\bar {\rm{p}}$p at
$\sqrt s$=1800 GeV
with the experimental points from the E710 measurement\cite{others}.
\begin{figure}[htb]
\centerline{\psfig{figure=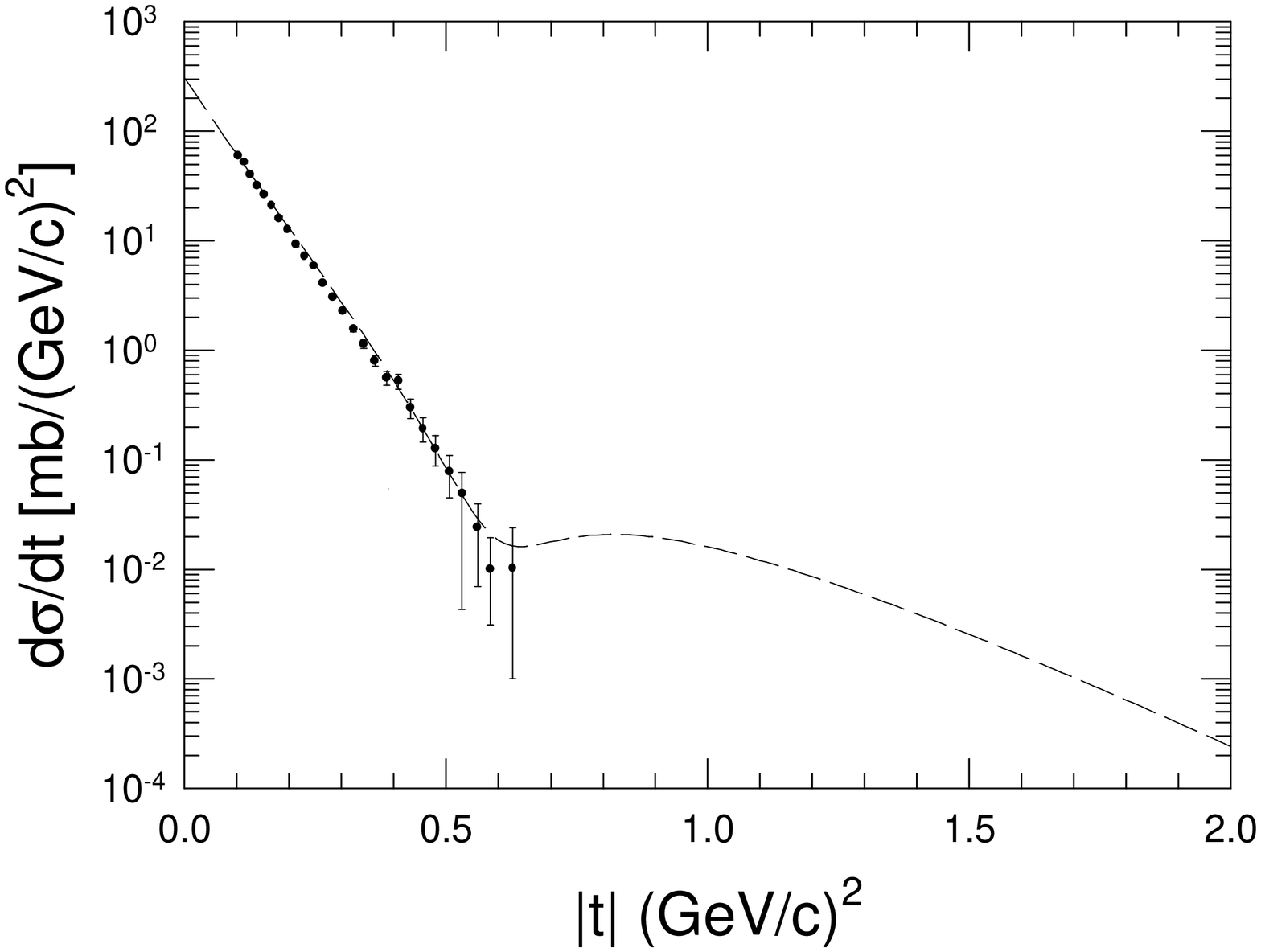,width=4.25in}}
\caption{\protect\small The differential elastic scattering cross section
 $\protect\frac{d\sigma}{dt}$ {\it vs.} $|t|$, for $\bar {\rm {p} }$p at
$\protect\sqrt s$=1800 GeV. The solid curve is the theoretical prediction based
on
a QCD model of soft interactions, and the experimental points are from the
E710 collaboration}
\end{figure}
There is good
agreement, both in shape and magnitude.  As we will see, this model
provides a good fit to current experimental data for the total cross section
$\stot$. This also implies that
the Tevatron energy, $\sqrt s$=1800 GeV, is on the
{\em threshold} of, but not in, `asymptopia'.
\section{An Eikonalized QCD Model}
We have
used a QCD-inspired, eikonalized model\cite{others} and found that
asymptotically\cite{others}, there is a critical impact parameter $b_c$ such
that
$\sigma_{\rm tot} = 2\pi b^2_c = 2 \pi \left( {J-1 \over \mu_{gg}} \right)^2
\log^2  {s\over s_0} \,.
$
The coefficient of the $\log^2 (s/s_0)$ term is given in terms of parameters
describing the gluon density of the nucleon, rather than the pion mass which
sets the scale of the $\log^2 s$ coefficient in the original Froissart bound.
The physical origin
of the rising cross section for a disk is seen to be the increasing
number of soft
gluons at small  $x$, where the gluon structure function behaves as
$x^{-J}$.
The large number of  gluons turns the proton into a disk with radius
$\mu^{-1}\simeq 0.8\,{\rm  (GeV)}^{-1}$. In  this model the eikonal behaves
{\em asymptotically} as
$s^{J-1}$. We find  $J-1\approx 0.05$--0.06, not
very  different from the power
behavior of $s^{0.086}$ as given by Donnachie and Landshoff's Pomeron
amplitude\cite{others}.

After including quarks as well as gluons,
contributions to the total cross section of a constant term, Regge type
terms and a $\log s$ term
also appear, which render the QCD model more complicated. They are however
essential when one attempts to apply these ideas to the transitional energy
regime of present experiments.
In general, the impact parameter amplitude $a(s,b)$ is
given by
$a(s,b)=\frac{i}{2}(1-e^{-\chi (s,b)}),$
where $b$ is the transverse distance in impact
parameter space, and $\chi (s,b)$ is the eikonal\cite{others}.
The nuclear amplitude $f_N(s,t)$ is given by
$f_N(s,t)=2\int b\;db\;J_0(b\sqrt{-t}\,)a(s,b),$
and the total cross section and the differential elastic
scattering cross section
are given by
$
\stot=4\pi \mbox{ Im }f_N(s,t), \,\,\, {\rm and}\,\,\,\,
\frac{d\sigma}{dt}=\pi|f_N(s,t)|^2, $
respectively.
For analyticity, we must have even and odd
eikonals $\chi_{even}(s,b)$
and $\chi_{odd}(s,b)$, where $\chi = \chi_{even} + \chi_{odd}$.
As $s\rightarrow\infty $,
$\chi_{odd}(s,b)$, which parameterizes the {\it {difference}}
between the $pp$ and
$p\bar{p}$ scattering amplitudes, vanishes.
The eikonal function $\chi$ is

$
2\chi (b,s)=W(b)\sigma(s)=P(b,s),$ where

$P=P_{gg}+P_{qg}+P_{qq}\,\,\,\,\, {\rm and} \,\,\,\,
W_{ij}(b)=\frac{\mu^2_{ii}}{96\pi}(\mu_{ii}b)^3 K_3 (\mu_{ii}b).$

\noindent We compute $P_{gg}$ exactly as
$P_{gg} (b,s)=W_{gg}(b) \sigma^{QCD}_{gg} (s),$
where, using $\tau =\hat{s}/s$,

$\sigma_{gg}^{QCD}(s)=\int d\tau
\,F_{gg}(x_1x_2\!=\!
\tau )\,\sigma_{gg}(\tau s),
$ and

$F_{gg}=\int dx_1 dx_2 \,f_g(x_1)\,f_g(x_2)\,\delta(x_1x_2-\tau).$

We use the gluon structure function
$f_g(x)\sim\frac{(1-x)^5}{x^J},$
with $J$, in Regge language, being the Pomeron intercept, and
$\sigma_{gg}(\hat{s})=\frac{9\pi\alpha_s^2}{\delta^2}\theta (\hat{s}-m_0^2).$
The high energy behavior is controlled by
\begin{eqnarray}
\lim_{s \rightarrow\infty}\mbox{ }\int^1_{m^2_0/s} d\tau F_{gg}(\tau)&
\sim&\int^1_{m^2_0/s} d\tau\frac{{}-\log\tau}{\tau^J}
\sim(\frac{s}{m^2_0})^{J-1},
\end{eqnarray}
which gives rise to the asymptotic energy dependence of the cross section
for the disk.
To reproduce the cross section at lower energies one must
consider the $qq$ and $qg$ contributions.  Using toy structure functions
$f_q\sim x^{-1/2}(1-x)^3$ and
$f_g\sim x^{-1}(1-x)^5$, we find
\begin{equation}
P_{qq}=W_{qq}(\mu_{qq}b)[ \frac{m_0}{\sqrt{s}}\log \frac{s}{s_0}+
{\cal P}(\frac{m_0}{\sqrt{s}})] \label{pqq1}
\end{equation}
and
\begin{equation}
P_{qg}=W_{qg}(\mu_{qg}b)
[a'\log\frac{s}{s_0}+{\cal P}'(\frac{m_0}{\sqrt{s}})].\label{pqg1}
\end{equation}
${\cal P}$ and ${\cal P}'$ are polynomials in $m_0/\sqrt{s}.$
Since we {\it{only}} attempt to decipher the high
energy behavior, we replace (\ref{pqq1}) and (\ref{pqg1}) by
\begin{eqnarray}
P_{qq}=W(\mu_{qq}b)[a+b\frac{m_0}{\sqrt{s}}]\,\,\,\,{\rm and}\,\,\,
P_{qg}=W(\sqrt{\mu_{qq}\mu_{gg}}b)[a'+b'\log \frac{s}{m^2_0}],\label{pqq2}
\end{eqnarray}
as suggested by standard Regge arguments.
Since the coefficients in Eqs.~(\ref{pqq2}) are treated as
free parameters,
this low energy parameterization can also accommodate elastic and
diffractive contributions.
We insure the correct analyticity properties
of our model amplitudes by substituting
 $
s\rightarrow~s~e^{-i\pi/2}
$ throughout.    \newline
We also introduce a crossing odd amplitude, suggested by Regge
theory, as

$
P_{odd}=W(\mu_{odd}b)a''\frac{m_0}{\sqrt{s}} e^{-i\pi/4}.
$

The data were simultaneously
fit for $\stot$, $\rho$ and $B$, for both pp and $\bar{\rm p}$p collisions,
in the energy region 15---1800 GeV (including the new Tevatron results and
the new UA4/2 $\rho$-value),
and the results are shown in
Figs. {\ref{sqcd}-\ref{bqcd}, respectively. As seen, the experimental data are
well reproduced
by the model, with a $\chi^2/d.f.=1.58$, with six fitted parameters.

High energy predictions from our QCD fits,
tightly bounded by the $\log s$  and Regge limits, are shown in
Table~{\ref{ta:predict}.
The quoted errors are
the statistical errors due to
the errors in the fitted parameters.

The Regge pole model\cite{others}, in which the scattering amplitude
grows as a power of $s$, \ie ,
$s^{0.086}$,
violates unitarity and thus can not be applicable in this form at
sufficiently high energies.
We regard the cross sections, $\stot\approx 115$\,mb at $\sqrt s=16$ TeV
and $\stot\approx 135$\,mb at $\sqrt s=40$ TeV, predicted by this
model as
upper limits, and are so listed in Table \ref{ta:predict}.

We see from Table \ref{ta:predict} that the central SSC value of 124 mb is
bracketed from above by 135 mb, which is
the Regge pole upper
limit, and from below by 117 mb, which is the extrapolation of an asymptotic
$\log s$ fit. Similarly, at the LHC energy,
the Regge pole upper limit of 115 mb and the $\log s$ lower limit of
104 mb tightly bracket our QCD prediction of 109 mb.\newpage
\begin{figure}[hb]
\centerline{\psfig{figure=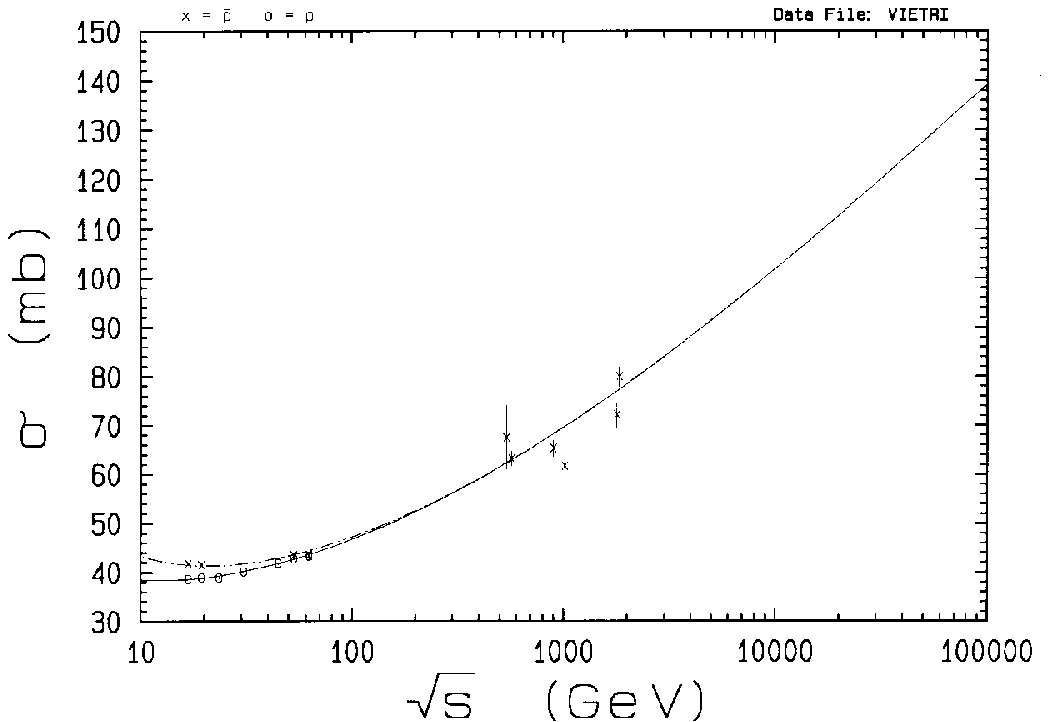,width=4.25in}}
\caption{$\stot$, the calculated and measured total cross sections in mb,
for $\bar{\rm p}$p
(dashed curve and crosses) and pp (full curve and circles), {\em vs.}
$\protect\sqrt s$, the energy, in GeV.\protect\label{sqcd}
}
\end{figure}
\begin{figure}[hb]
\centerline{\psfig{figure=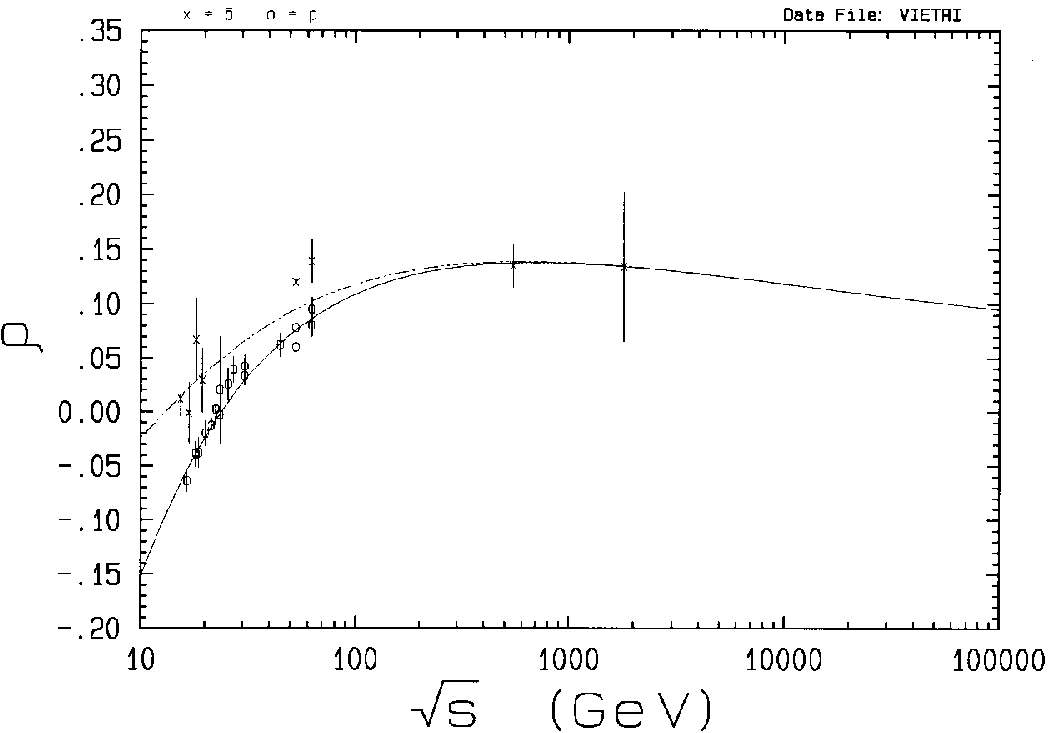,width=4.25in}}
\caption{$\rho$, the calculated and measured ratios
${\rm Re}f(0)/{\rm Im}f(0)$, for $\bar{\rm p}$p
(dashed curve and crosses) and pp (full curve and circles), {\em vs.}
$\protect\sqrt s$, the energy, in GeV.\protect\label{rqcd}
}
\end{figure}\newpage
\begin{figure}[hb]
\centerline{\psfig{figure=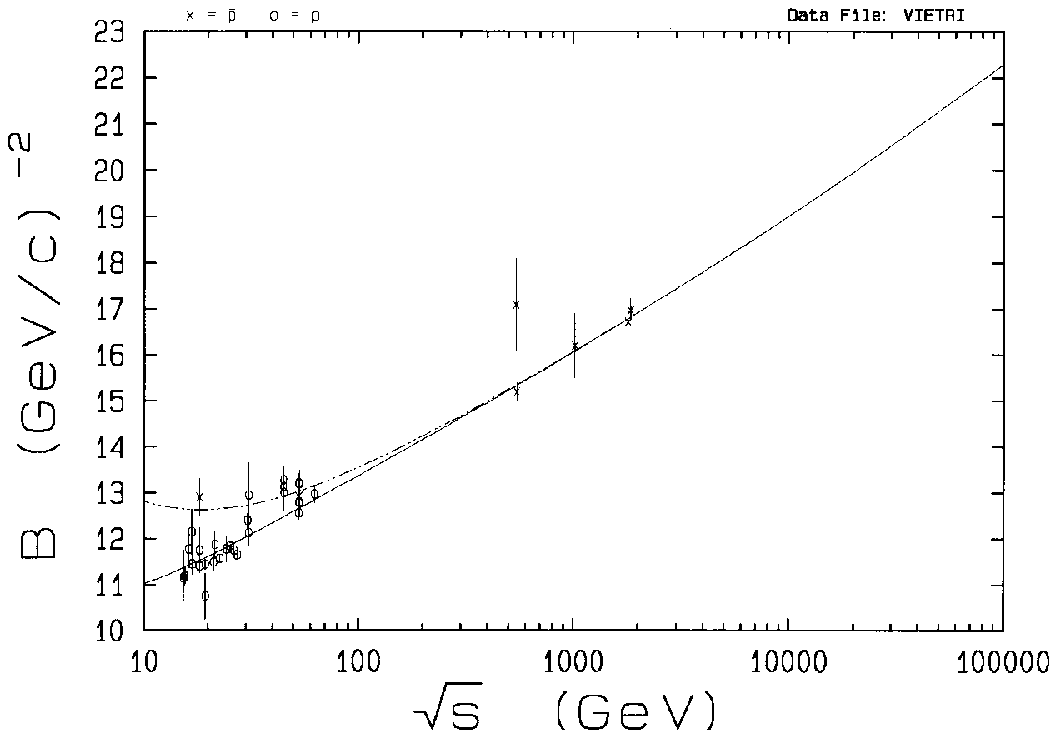,width=4.25in}}
\caption{$B$, the calculated and measured nuclear slope parameters in
${\rm (GeV/c)}^{-2}$,
for $\bar{\rm p}$p
(dashed curve and crosses) and pp (full curve and circles) {\em vs.}
$\protect\sqrt s$, the energy, in GeV.\protect\label{bqcd}
}
\end{figure}
\begin{table}[hb]  
%
\caption{{\protect\small Collider cross section predictions, with upper
and lower bounds,
using `QCD'.}
\label{ta:predict}}
$$          
\BeginTable
     \BeginFormat
 |  c            |         c      |   c          | c         |   c     |
     \EndFormat
 "{}& \use4 \= &\\0
 "{}| $\sqrt{s}$ | Lower Bound ($\log{s}$) \|* QCD (Eikonal) \|* Upper Bound
 (Regge)|\\+20
 " {} | (TeV)    | (mb)                    \|* (mb)          \|* (mb) |\\+02
 &\use5 \= &\\0
 |LHC " 16       | $104\pm 1$            \|* $109\pm 4$    \|* $115$ |\\+40
 &\use5 \=&\\0
 |SSC " 40       | $117\pm 1$            \|* $124\pm 4$    \|* $135$ |\\+04
 & \use5 \= & \\0
\EndTable
$$
\end{table}

\section{Conclusions}
We note that in actual numerical fits, these two very
different models give virtually indistinguishable
results in the energy region in which data are available, a
similarity which {\em must disappear} at
sufficiently high energies. Although both models easily
accommodate the data at
present energies,
the physics of each model is significantly different.  Most importantly,
these two models ascribe
the rise of the total cross section as due to
\begin{romList}
\item \Label{asym} an asymptotic term, which
(the data will show) behaves as
$\log \,(s/s_0)$, where $s_0$
is a scale for $s$, for Model 1,
\item \Label{partons} the dramatic increase
of the number of soft partons, for Model 2,
\end{romList}
respectively.

We conclude:
\begin{romList}
\item there is {\em no experimental evidence} from forward scattering
parameters for the existence of  `odderons', \ie, odd amplitudes that
do not vanish with increasing energy.
\item the data {\em at present energies} indicate that the total cross
section
is rising as
$\log \,(s/s_0)$, and {\em not} as $\log^2 (s/s_0)$, which is probably an
indicator that we are not yet in `asymptopia'.
\end{romList}

Our QCD fit predicts
$\stot(16\,\, {\rm  TeV})=109\pm 4\,$mb and
 $\stot(40\,\, {\rm  TeV})=124\pm 4\,$mb, results which are tightly
bracketed from {\em above} by Regge predictions and from {\em below} by a
$\log (s)$ fit.

\end{document}